\documentclass[a4paper,fleqn,usenatbib,useAMS]{mnras}

\usepackage{graphicx}	
\usepackage{amsmath}	
\usepackage{amssymb}	
\usepackage{multicol}        
\usepackage{bm}		
\usepackage{pdflscape}	

\usepackage[normalem]{ulem}

\usepackage[T1]{fontenc}
\usepackage{ae,aecompl}

\usepackage{newtxtext,newtxmath}
\usepackage{hyperref}

\title[Comet C/2016 R2]{\textit{Ionic emissions in comet C/2016 R2 (Pan-STARRS) }}

\author[Venkataramani et al.]{Kumar Venkataramani$^{1,2}$\thanks{E-mail : kumar.venkataramani@gmail.com} 
Shashikiran Ganesh$^{1}$ and Kiran S.Baliyan$^{1}$
\\ \\
$^{1}$ Astronomy \& Astrophysics Division, Physical Research Laboratory, Ahmedabad, India. \\
$^{2}$ Department of Physics, Leach Science Center, Auburn University, Auburn, AL, USA.}

\pubyear{2019}

\begin{document}
\newcommand\Tstrut{\rule{0pt}{2.6ex}}
\newcommand\Bstrut{\rule[-0.9ex]{0pt}{0pt}}

\label{firstpage}
\pagerange{\pageref{firstpage}--\pageref{lastpage}}
\maketitle

\begin{abstract}
We carried out observations of a peculiar comet, C/2016 R2 (Pan-STARRS), using a low resolution spectrograph mounted on the 1.2m telescope at Mount Abu Infrared Observatory, India. The comet was observed on two dates in January 2018, when it was at a heliocentric distance of 2.8 AU. 
Study based on our observations revealed that the optical spectrum of this comet is quite unusual as compared to general cometary spectra. Most of the major cometary emissions like C$_{2}$, C$_{3}$ and CN were absent in comet C/2016 R2. However, the comet spectrum showed very strong emission bands from ionic species like CO$^{+}$ and N$_{2}^{+}$. A mean N$_{2}$/CO ratio of 0.09 $\pm$ 0.02 was derived from the spectra and an extremely low depletion factor of 1.6 $\pm$ 0.4 has been estimated for this ratio as compared to the solar nebula. We have also detected minor emission features beyond 5400 \AA, albeit marginally. The column densities of CO$^{+}$ and N$_{2}^{+}$ were calculated from their emission bands.
The optical spectrum suggests that the cometary ice is dominated by CO. The low depletion factor of N$_{2}$/CO ratio in this comet, as compared to the solar nebula and the unusual spectrum of the comet are consequences of distinctive processing at the location of its formation in the early solar nebula.
\end{abstract}


\section{Introduction}
Cometary molecular emissions are well known and have been studied since a long time. A typical optical spectrum of a comet with well developed coma shows molecular emissions \citep{c2014q2} dominated by carbon chain radicals like C$_{2}$ and C$_{3}$. NH$_{2}$ and CN are two other species which show prominent emission lines in the optical spectrum. 
In general, these molecular emissions start appearing sequentially when the comet comes closer than 3 AU \citep{comets_book_ks} to the Sun. The most likely emission to appear first is that of CN molecule at around 3 AU, followed by the rest of the emissions. There are very few comets in which emissions are reported beyond 3 AU and even fewer beyond 5 AU.
Ionic emissions like CO$^{+}$ and N$_{2}^{+}$ are rarely seen in the coma of a comet. They are however abundantly found in the plasma tail of comets. 
Comet 1962 VIII, also known as comet Humason, is one where CO${+}$ emissions were detected\citep{comet_humason} in the coma.
Lot of work \citep[e.g.][]{ comet_halley_co_1,comet_halley_co_2, comet_halley_co_3,comet_halley_co_4} has been carried out based on the CO$^{+}$ emissions that were found in the coma and tail of Comet 1P/Halley. 
There are several more comets in which significant amount of CO$^{+}$ has been detected e.g. 109P/Swift-Tuttle \citep{swift-tuttle}, 122P/1995 S1 (deVico) \& C/1995 O1 (Hale-Bopp) \citep{hale-bopp-cochran}, comet C/1996 B2(Hyakutake) \citep{hyakutake_1} to name a few. CO is the presumed dominant parent source for the production of CO$^{+}$ ions in comets, although dissociative ionization of CO$_{2}$ also produces CO$^{+}$ ions in comets \citep{swift-tuttle-2}. 
CO has been detected in all of the comets showing CO$^{+}$ emissions. In fact, 
strong (A-X) transition CO bands in the UV region (near 1500 \AA $ $) have been detected in all the bright comets observed with IUE or HST \citep{comets_book_ks}. \\

Molecular nitrogen (N$_{2}$) is also one of the most abundant species in the star-forming regions \citep{n2_star_1} and in the proto-solar nebulae\citep{n2_solar_1, n2_solar_2}. However, there have been a lot of speculations about its presence in comets, even though N$_{2}^{+}$ has been observed in many comets with mostly low resolution spectra. \citet{n2+_comets_1} claim to have observed N$_{2}^{+}$ at a record heliocentric distance of 6.8 AU. \citet{hale-bopp-cochran} \& \citet{n2+_comets_3} have reported N$_{2}^{+}$ in comets 122P/de Vico and C/2002 C1 (IKEYA-ZHANG) respectively, whereas \citet{n2+_comets_4} claims to have observed this emission band in comet C/1995 O1 (Hale-Bopp). A strong confirmation for the presence of N$_{2}$ in comets came from the in-situ measurements of comet 67P/C-G made by ROSINA mass spectrometer on board the Rosetta spacecraft \citep{rosetta_n2_co_1}. The N$_{2}$/CO ratio in comet 67P/ determined by \citet{rosetta_n2_co_1} is probably the most authentic value reported, as it was measured in-situ. \citet{rosetta_n2_co_2} have compared this N$_{2}$/CO ratio in comet 67P with various laboratory results and have discussed the formation, agglomeration and origin of these ices on comets. 

Comet C/2016 R2 (`R2' for all future text references in this paper) is an Oort cloud comet discovered by PanSTARRS on September 7$^{th}$ 2016. The comet brightened to a magnitude of 13 in visual band in January 2018. The comet reached its perihelion in May 2018. It has an orbital eccentricity of 0.996, semi-major axis of 738 AU, orbital inclination of 58 deg to the ecliptic and a total orbital period of 20084 years. Initially, \citet{c2016r2_atel} reported the observations of CO (2-1) rotational lines in this comet. \citet{cochran_co_c2016r2, cochran_co_c2016r2_erratum} have detected strong CO$^{+}$ and N$_{2}^{+}$ emission lines in comet R2 and have estimated the N$_{2}$/CO ratio. However, they have not detected any emissions from the neutral species which dominate the optical spectrum of a comet. Observations carried out by \citet{comet_r2_CO} in the submillimeter wavelengths have indicated that the comet is rich in CO and depleted in HCN. \citet{n2_co_Biver} have also confirmed the lack of HCN in this comet. More recent reports have established the uniqueness of this comet. \citet{n2_co_Mckay} have discussed in detail, the peculiar composition of R2 using observations from various ground and space observing facilities. \citet{n2_co_opitom} have obtained high resolution optical spectra of R2. They claim the detection of various ionic species and very faint emissions from the neutrals. Detection of major emissions from some of the neutral species is expected at such heliocentric distances and their non detection underlines the need for a detailed study of this comet. In his thesis work, \citet{Kumar_thesis} has observed and studied this comet in detail using the low resolution spectrograph LISA.
Here we investigate the optical spectra of R2 by observing it with a low resolution spectrograph. We derive and present the column densities of the cometary emissions based on our observations. We also discuss the implications of the N$_{2}$/CO ratio determined for comet R2 from our observations.

\section{Observation and Data Analysis}
The observations of comet R2 were made using a low resolution spectrograph called LISA\footnote{More details on the spectrograph are available on the web site of the manufacturer: Shelyak Instruments (\url{http://www.shelyak.com/})} mounted on the 1.2m telescope at the Mount Abu Infrared observatory, India. The spectrograph covers a wavelength range of 3800 \AA~ to 7400 \AA. The 35$\mu$m slit used in the spectrograph results in a wavelength dispersion of 2.6 \AA~ per pixel. The observations are described in detail by \citet{c2014q2}. On the 1.2m telescope, the plate scale on the LISA spectroscopic detector plane is 0.25 arcsec per pixel. The slit has a width of 1.75 arcsec and is 98 arcsec long. The slit was placed along the North South Direction.

\begin{table*}
\centering
\caption{Observational Log}
\begin{tabular}{|c | c  | c | c | c | c | c |}
\hline 
\Tstrut\Bstrut Date & Mid-UT & Heliocentric   & Geocentric  & Solar Phase  & Exposure & Airmass \\
\Tstrut\Bstrut     &        & Distance r$_{h}$ & Distance $\Delta$ &        &          &          \\
\Tstrut\Bstrut     &        &  (AU)            & (AU)              &  (degrees) &  (Seconds)  &    \\ \hline
\Tstrut\Bstrut 13/01/2018 & 15:55 & 2.87 & 2.13 & 15 & 1800 & 1.00 \\ \hline

\Tstrut\Bstrut 25/01/2018 & 14:30 & 2.82 & 2.23 & 18 & 1800 & 1.00\\

\hline
\end{tabular}
\label{comet_obs}
\end{table*}

The spectra of the comet were obtained on two nights, i.e. 2018 January 13 and 2018 January 25. The observation log is shown in table \ref{comet_obs}. On both the epochs, the comet was observed at an elevation of 83 deg (close to zenith), at an airmass $\approx$ 1.  The sky conditions were photometric on both the nights. The comet was observed at a relatively low solar phase angle of 15 to 18 deg. Hence, though the slit was placed at comet's photo-centre, the geometry of observation was such that the comet's tail might be in the direction along the line of sight. As a consequence, we are integrating the flux coming from all along the tail of the comet. In order to subtract the contribution from the sky, the telescope was moved away from the comet by nearly 30 arc minutes and sky spectra were taken with the same exposure times.  The following flux standards, available in the IRAF standard star database, were observed for the purpose of flux calibration : HD74721, HD81809, HD84937, BD+08d2015. Spectral extraction and data analysis were carried out using IRAF and GNU Octave software respectively.

\section{Results}

\begin{figure*}
\centering
  \includegraphics[width=1\textwidth]{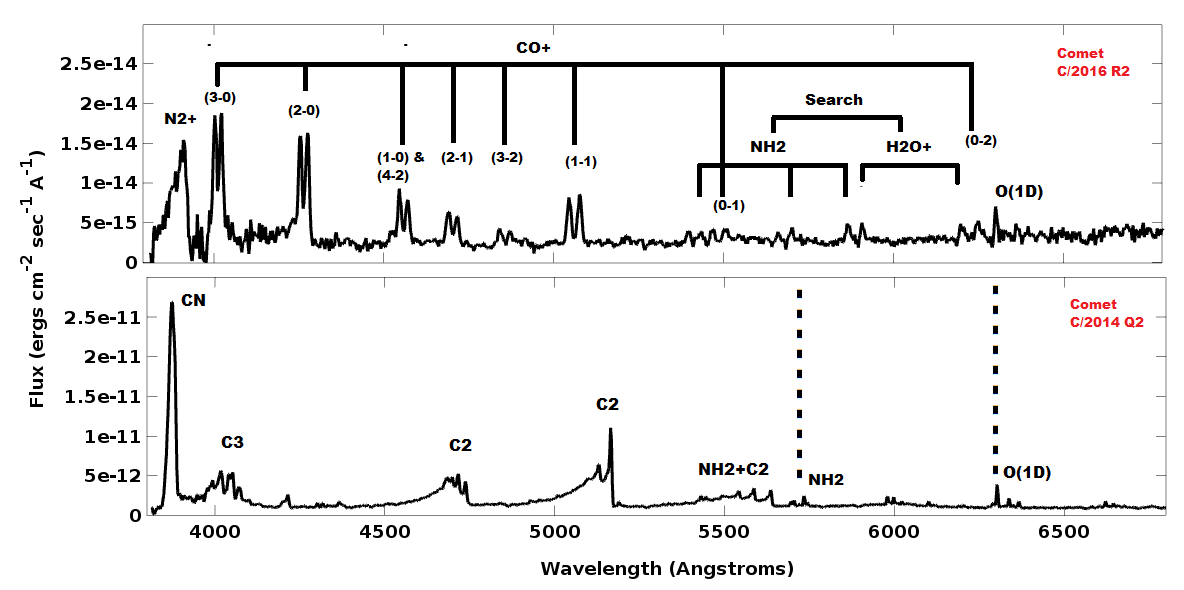}
  \caption{Top: The optical spectrum of comet C/2016 R2 taken on 25th January 2018. The CO$^{+}$ emission lines are marked. The emission features beyond 5400 \AA~ have also been marked. These features were investigated to search for possible emissions from H$_{2}$O$^{+}$ and NH$_{2}$. Bottom: The spectrum of comet C/2014 Q2 \citep{c2014q2} been shown in order to compare the two spectra. This shows emissions from carbon chain radicals like C$_{2}$, C$_{3}$. CN and resembles a spectrum of a typical comet. Comet R2 does not show any of these lines and its spectrum is dominated by CO$^{+}$ emissions.}
 \label{spectra}
\end{figure*}

The optical spectrum of the comet R2 was obtained with the long slit of the spectrograph centred on the coma of the comet. The spectrum obtained on 25th January 2018 is shown in the figure \ref{spectra}. The major emission features are marked in the spectrum and are listed in table \ref{features}. 
The major cometary emissions such as the C$_{2}$ Swan bands, C$_{3}$ and CN are completely absent in the spectrum of comet R2. Even though the carbon chain molecules (C$_{2}$ and C$_{3}$) are expected to be seen only at much closer heliocentric distances, as compared to the Sun-comet distance at the time of these observations, the least presumption would be to expect at least CN emissions at this distance. However, no such emissions were detected. The lack of C$_{2}$, C$_{3}$ and CN in our observed spectrum is consistent with results reported from observations before \citep{cochran_co_c2016r2, cochran_co_c2016r2_erratum} and after \citep{n2_co_opitom} our observation dates.  The optical spectrum on the contrary was dominated by a number of CO$^{+}$ emission bands (comet tail system). With low resolution spectra, it is usually difficult to resolve and differentiate CO$^{+}$ bands from the major emission features of C$_{2}$, C$_{3}$ and CN. However, due to the absence of such emission bands from the neutral species in comet R2, we were easily able to detect most of the CO$^{+}$ bands in the region of 4000 \AA~ to 5500 \AA.
Along with CO$^{+}$, we have also detected a strong emission band feature with a band head centred around 3910 \AA $ $ which is attributed to N$_{2}^{+}$ emission. In addition to the emissions detected by \citet{cochran_co_c2016r2, cochran_co_c2016r2_erratum} with a high resolution spectrograph, we also report the detection of NH$_{2}$ bands (0-10-0, 0-11-0, 1-7-0 being the most prominent among them) and two significant emission features around 5910 \AA~ and 6200 \AA, which could possibly be attributed to H$_{2}$O$^{+}$ ions. Moreover, we have observed the comet at a heliocentric distance of 2.8 AU in January 2018, whereas \citet{cochran_co_c2016r2, cochran_co_c2016r2_erratum} have observed it at a much larger distance of 3.09 AU. These observations are critical as many of the major cometary emissions are triggered at these distances. 

As the CO$^{+}$ emissions were extremely strong, the bands were identified by matching the wavelengths of the observed emission peaks with the $\Pi_{1/2}$ and $\Pi_{3/2}$ band head emissions as mentioned in \citet{co_theory} (Table 3 of the paper). However, in order to identify the weaker emissions beyond 5400 \AA, cometary lines listed in the catalogues (described in section \ref{line_id}) and lab spectra have been used. We have obtained the column densities of these species seen in the coma of the comet at two dates in January 2018.    

\begin{table}
\centering
\caption{ Emission band features detected in the spectrum of comet R2. The wavelength range specifies the total extent of the band. The NH$_{2}$ and H$_{2}$O$^{+}$ bands mentioned in the table indicates that a search for emissions from these species have been carried out in the corresponding wavelength ranges.}
\begin{tabular}{|c | c | c | c |}
\hline
Wavelength \Tstrut\Bstrut & Molecular & Band & Ref \\ 
\Tstrut\Bstrut (\AA)      & Species &        &    \\ \hline
 \Tstrut\Bstrut 3978-4048 & CO$^{+}$ & (3-0) & 1 \\ \hline 
 \Tstrut\Bstrut 4219-4292 & CO$^{+}$ & (2-0) & 1  \\ \hline
 \Tstrut\Bstrut 4663-4739 & CO$^{+}$ & (2-1) & 1 \\ \hline
 \Tstrut\Bstrut 4803-4888 & CO$^{+}$ & (3-2) & 1 \\ \hline
 \Tstrut\Bstrut 5017-5092 & CO$^{+}$ & (1-1) & 1 \\ \hline
 \Tstrut\Bstrut 4502-4551 & CO$^{+}$ & (4-2) & 1  \\ \hline
 \Tstrut\Bstrut 4529-4581 & CO$^{+}$ & (1-0) & 1  \\ \hline
 \Tstrut\Bstrut 5450-5522 & CO$^{+}$ & (0-1) & 1  \\ 
 \Tstrut\Bstrut           &          & (blend with  &  \\   
 \Tstrut\Bstrut           &          & NH$_{2}$ 0-11-0) &  \\ \hline   
 \Tstrut\Bstrut 4529-4581 & CO$^{+}$ & (0-2) & 1  \\
 \Tstrut\Bstrut           &          & (blend with &  \\
 \Tstrut\Bstrut           &          & H$_{2}$O$^{+}$ 0-8-0) &  \\ \hline
 \Tstrut\Bstrut 5672-5712$^{*}$ & NH$_{2}$ & 0-10-0 & 2,3  \\ \hline
 \Tstrut\Bstrut 5410-5495 & NH$_{2}$ & 0-11-0 & 2,3  \\ \hline
 \Tstrut\Bstrut 3810-3935 & N$_{2}^{+}$ & 0-0 & 4   \\ \hline
 \Tstrut\Bstrut 5893-5916 & H$_{2}$O$^{+}$ & 0-9-0 & 3   \\ \hline
 \Tstrut\Bstrut 6183-6216$^{*}$ & H$_{2}$O$^{+}$ & 0-8-0  & 3   \\ 
 \Tstrut\Bstrut           &          & (blend with   &  \\ 
 \Tstrut\Bstrut           &          & CO$^{+}$ (0-2) band )  &  \\ \hline 
 \end{tabular}
 \\ \vspace*{0.1in} 1-\citet{co_theory}, 2-\citet{nh2_g_factor}, 3-\citet{cometary_lines}, 4-\citet{n2_g_factor}
\label{features}
\small{** Contaminated with CO$^{+}$ doublet}

\end{table}

\subsection{CO$^{+}$ emission bands}
Many strong CO$^{+}$ emission bands were detected in the optical spectrum of the comet (listed in table \ref{features}).  All these emission bands belong to the comet-tail system of the CO$^{+}$ transitions. \citet{co_theory} have given a theoretical estimate of the relative band intensities of the CO$^{+}$ emission bands. They have compared their results with those on comet Humason or 1962 VIII \citep{comet_humason}. We have calculated the band intensities of CO$^{+}$ emission bands in comet R2 at a heliocentric distance of 2.8 AU. 

\begin{table*}
\caption{ Relative intensities of CO$^{+}$ emission bands (Comet-tail system). All the band intensities are normalized to the (3-0) band}
\begin{tabular}{|c | c | c | c | c | c |}
\hline
Band Transition \Tstrut\Bstrut & \multicolumn{2}{|c|}{This work (Comet C/2016 R2)} & Arpigny(observations)$^1$ & A'Hearn (Theory)$^2$ & Krishnaswamy (Theory)$^3$ \\
\cline{2-3}
      & 13/01/2018 \Tstrut\Bstrut & 25/01/2018 & Comet Humason  & &  \\ \cline{1-6}
(2-0) \Tstrut\Bstrut & 0.77 & 0.88 & 0.86 & 0.83 & 0.99 \\ \hline 
(2-1) \Tstrut\Bstrut & 0.25 & 0.27 & 0.37 & 0.41 & 0.38 \\ \hline
(3-2) \Tstrut\Bstrut & 0.10 & 0.12 & 0.21 & 0.21 & 0.18 \\ \hline
(1-1) \Tstrut\Bstrut & 0.32 & 0.37 &  -   &  -   &  -    \\ \hline
(4-2) \Tstrut\Bstrut & 0.22 & 0.17 & 0.22 & 0.21 & 0.16  \\ \hline
(1-0) \Tstrut\Bstrut & 0.32 & 0.36 & 0.51 & 0.52 & 0.52  \\ \hline
\end{tabular}
\\
\label{band_intensities}
$^1$ \citet{comet_humason}; $^2$ \citet{co_theory};  $^3$\citet{co_ratio_KS}\\
\end{table*}

Comet Humason was observed at around 2.6 AU and the theoretical values are calculated for 1 AU sun-comet distance. Two of the strongest CO$^{+}$ emission bands seen in the spectrum are due to the (2-0) and (3-0) A-X transition. However, the (2-0) band is contaminated by (0-1) N$_{2}^{+}$ emission (Described in section \ref{n2_band_section}). Therefore, the band intensities are normalized to (3-0) band and these are shown in table \ref{band_intensities}. The values reported in other works have also been listed in the table alongside our calculated values. (All the values are normalized to the (3-0) band). 

\begin{table*}
\centering
\caption{ CO$^{+}$ emission bands in comet C/2016 R2 }
\label{co_production}
\begin{tabular}{|c | c | c | c | c |}
\hline
Band Transition \Tstrut\Bstrut & Wavelength range & g-factor & \multicolumn{2}{c|}{Column Density}  \\
     &  (\AA)      &  ($\times$ 10$^{-14}$ ergs mol$^{-1}$ sec$^{-1}$) & \multicolumn{2}{c|}{($\times$ 10$^{12}$ mol cm$^{-2}$)} \\
\cline{4-5}
     \Tstrut\Bstrut & & &   13/01/2018  & 25/01/2018 \\ \cline{1-5}
(3-0) \Tstrut\Bstrut & 3978-4048 & 2.20 & 2.20 & 1.88  \\ \hline 
(2-0) \Tstrut\Bstrut & 4219-4292 & 1.79 & $<$ 2.08 & $<$ 1.78 \\ \hline
(2-1) \Tstrut\Bstrut & 4663-4739 & 0.89 & $<$ 1.36 & $<$ 1.16  \\ \hline
(3-2) \Tstrut\Bstrut & 4803-4888 & 0.45 & 1.06 & 0.91  \\ \hline
(1-1) \Tstrut\Bstrut & 5017-5092 & 1.27 & 1.23 & 1.05  \\ \hline
(4-2) \Tstrut\Bstrut & 4502-4551 & 0.44 & 2.46 & 2.11  \\ \hline
(1-0) \Tstrut\Bstrut & 4529-4581 & 1.03 & 1.52 & 1.30  \\ \hline

\end{tabular}
\label{co_production}
\end{table*}

The values of the relative band intensities for comet R2 in table \ref{band_intensities} match quite well, on both epochs, with the other values reported. The emission band fluxes on both epochs were converted to column densities using the equation 4 of \citet{c2014q2}. These are listed in table \ref{co_production}. The g-factors given by \citet{co_theory} have been used for the calculations.

\subsection{N$_{2}^{+}$ band}
\label{n2_band_section}
A very strong emission band of N$_{2}^{+}$, with band-head centred around 3910 \AA~ was detected in our spectrum of comet R2. This emission comes from the (0-0) band transition of the first-negative system of N$_{2}^{+}$ ion. Major difficulty in detecting the N$_{2}^{+}$ bands in this wavelength region is caused by the blending of this band with the strong CN emission which generally over-shadows any minor N$_{2}^{+}$ emission which might be present. By comparing the observed spectra of comet R2 with that of a typical optical cometary spectra like that of comet C/2014 Q2 \citep{c2014q2}, we can clearly see that the CN emission band is missing in this spectrum. Due to the absence of CN emission in comet R2, there was no contamination present in the observed N$_{2}^{+}$ band. The cometary origin of this band has been well justified by \citet{cochran_co_c2016r2, cochran_co_c2016r2_erratum}. In our case, although the observations were made in the early half of the night, we expect that the contamination due to the terrestrial N$_{2}^{+}$ skylines would be very low since the N$_{2}^{+}$ emissions from the sky are extremely weak during the night \citep{n2_nightglow}    
\begin{figure*}
\centering
  \includegraphics[width=1\textwidth]{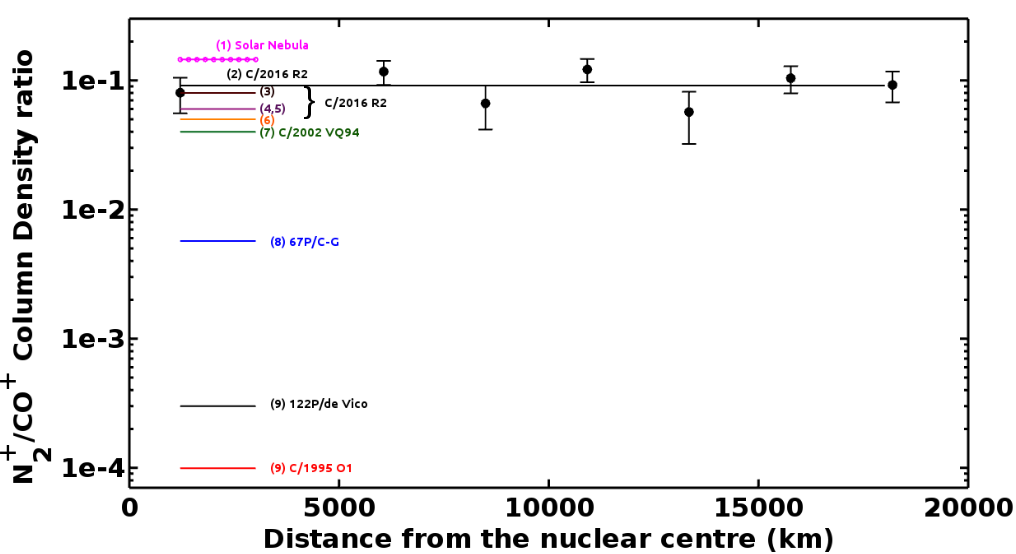}
   \caption{The column density ratio of N$_{2}^{+}$ to CO$^{+}$ as a function of distance from the photo-centre. All the ratios have been determined with respect to the CO$^{+}$ (3-0) band. The N$_{2}$/CO ratio of the solar nebula and for a few other comets have been marked in the figure, for comparison. The line shown for comet R2 is the mean value of the ratio determined at different distances from the photo-centre. For the other comets, the horizontal lines represent the reported value for respective objects independent of the distance. 
   The numbers listed in the figure (same as in Table \ref{n2_co_ratios}) correspond to the following references:  1 \citet{solar_nebula_value}, \citet{rosetta_n2_co_1}; 2. This Work/\citep{kumar_epsc,Kumar_thesis}; 3. \citet{n2_co_Biver}; 4. \citet{cochran_co_c2016r2, cochran_co_c2016r2_erratum};  5. \citet{n2_co_opitom}; 6. \citet{n2_co_Mckay}; 7. \citet{n2+_comets_1};  8. \citet{rosetta_n2_co_1}; 9. \citet{hale-bopp-cochran} 
   }
 \label{ratio}
\end{figure*}
and any minimal emission would have been subtracted in the process of sky subtraction. 

The wavelengths of the band heads for different band transitions of N$_{2}^{+}$ and their relative strengths have been tabulated by \citet{N2+_line_list}. The (0-1) and (0-2) band transitions of N$_{2}^{+}$ ion give rise to emissions with band heads centred around 4278 \AA~ and       \AA~ respectively. These emissions, if present would get strongly blended with the CO$^{+}$ (2-0) and (2-1) emission bands respectively. However, their strengths are extremely weak as compared to the stronger (0-0) band. According to the relative strengths given in table 57 of \citet{N2+_line_list}, the (0-1) and (0-2) bands are around 25\% and 3\% as intense as the (0-0) band, respectively. With this, we estimate the contamination of CO$^{+}$ (2-0) band by the corresponding N2+ bands to be about 27\% and that of (2-1) band to be about 11\%.  

The g-factor given by \citet{n2_g_factor} has been used to calculate the column densities of the N$_{2}^{+}$ band. The abundance ratios of N$_{2}$/CO play an important role in understanding the early solar nebula. We presume that the major source of N$_{2}^{+}$ and CO$^{+}$ is the photo-ionization of N$_{2}$ and CO respectively.  An indirect method to calculate the abundance ratios is to estimate the ratio of column densities of the corresponding observable ion species. 
Therefore, we have obtained the column density ratios of N$_{2}^{+}$/CO$^{+}$ from the spectrum of the comet R2. \citet{cochran_co_c2016r2,cochran_co_c2016r2_erratum} have reported a value of 0.06 for the N$_{2}^{+}$/CO$^{+}$ column density ratio. They have used the CO$^{+}$ (2-0) band flux for calculating the ratio. However, due to the contamination of CO$^{+}$ (2-0) band by N$_{2}^{+}$ (0-1) band, we are forced to use the integrated flux of CO$^{+}$ (3-0) band to estimate this ratio.  We sum the entire flux in the CO$^{+}$ (3-0) band in order to calculate the ratios. Taking advantage of the long slit, we have obtained the N$_{2}^{+}$/CO$^{+}$ column density ratios at different distances from the photo-centre as mapped along the slit. Although the slit covers a total of 98 arcsec on the sky, the coma profile could only be measured upto 12.7 arcsec on either side of the photo-centre. This translates to a distance of 18195 km from the centre. Beyond this distance, SNR was too low to be measured. The N$_{2}^{+}$/CO$^{+}$ column density ratio remains fairly constant within the distance that we have covered. We obtain a mean value of 0.09 $\pm$ 0.02 on both the dates of observation in January 2018, when the comet was at a heliocentric distance of 2.8 AU \citep[Initial work reported in][]{kumar_epsc, Kumar_thesis}.
At this point, it would be appropriate to discuss the issue of N$_{2}^{+}$ (0-0) band being contaminated by the CO$^{+}$ (5-1) emission, while determining N$_{2}^{+}$/CO$^{+}$ ratio. Such contamination would lead to overestimation of this ratio. We, therefore state that the reported value is not an exact one, but is an upper cut-off value. This could possibly explain the larger value of this ratio, as compared to the one reported by \citet{cochran_co_c2016r2, cochran_co_c2016r2_erratum}. Since the relative strength of the CO$^{+}$ (5-1) band is not known (it has not been explicitly determined by \citet{co_theory}), we assume, that its contribution is negligible. If it were not the case, our value would have differed significantly from the value reported by \citet{cochran_co_c2016r2, cochran_co_c2016r2_erratum} obtained with high resolution spectra. It is also important to note that \citet{n2_co_Biver} have reported a mean value of 0.08 for the N$_{2}^{+}$/CO$^{+}$ column density ratio. This is close to the value measured by us and is within our error limits. In fact, \citet{n2_co_Biver} report a value of 0.09 $\pm$ 0.03 for the ratio measured on 22nd December 2017 with their LISA spectrograph, which is consistent with the value measured by us. For all further discussions in this paper, N$_{2}^{+}$/CO$^{+}$ ratio is taken to be as measured by us. \\
The conversion of ion abundance ratio to the neutral abundance ratio N$_{2}$/CO, requires the knowledge of the reaction rates. Assuming that photo-dissociation of N$_{2}$ and CO by solar UV flux is the major contributor for N$_{2}^{+}$ and CO$^{+}$ ions respectively and that the electron dissociative recombination is the primary loss mechanism for both ions \citet{n2_nh3_ratio}, the conversion factor from N$_{2}^{+}$/CO$^{+}$ to N$_{2}$/CO is calculated using the reaction rates used by \citet{n2_nh3_ratio}. The conversion factor turns out to be 0.98. Since the uncertainty in the measured ratio is greater than 20$\%$, the conversion factor would be indistinguishable from unity. Therefore we assume that N$_{2}^{+}$/CO$^{+}$ $\approx$ N$_{2}$/CO. The column density and the column density ratios have been listed in table \ref{n2_production}.

\newcommand\TTstrut{\rule{0pt}{3ex}}
\newcommand\BBstrut{\rule[-1ex]{0pt}{0pt}}

\begin{table*}
\centering
\caption{ N$_{2}^{+}$ emission band [B-X(0-0) transition] in comet C/2016 R2 - Wavelength range : 3810-3935 \AA }
\begin{tabular}{|c | c | c |}
\hline
Date \Tstrut\Bstrut & Column Density &  Column Density Ratio  \\ \hline
\TTstrut\BBstrut     &  ($\times$ 10$^{11}$ mol cm$^{-2}$) & \textbf{$\frac{N(N_{2}^{+})}{N(CO^{+})}$}  \\ \hline
\Tstrut\Bstrut 13/01/2018 &  2.42   &   0.09   \\ \hline
\Tstrut\Bstrut 25/01/2018 &  2.55   &   0.09  \\ \hline
\end{tabular}
\label{n2_production}
\end{table*}

The N$_{2}$/CO ratio is of great significance as it helps to understand the formation processes in comets and in the early solar nebula. The significance and implications of this ratio have been discussed at length in section \ref{discussion}.
The plot of this ratio as a function of the radial distance from photo-centre is shown in figure \ref{ratio}. The values of N$_{2}$/CO ratio for a few other comets and for the solar nebula have also been marked, for comparison. The N$_{2}$/CO ratio for comet R2 obtained from various observations \citep{cochran_co_c2016r2, cochran_co_c2016r2_erratum, n2_co_Biver, n2_co_Mckay, n2_co_opitom} and from this work, along with the value of this ratio measured for other comets  have been tabulated in table \ref{n2_co_ratios}. \\

\begin{table*}
\centering
\caption{ The N$_{2}$/CO ratio for the Solar nebula, various comets and comet R2 obtained from various observations. The numbers listed in the final column (same as in Fig. \ref{ratio}) correspond to the following references:  1 \citet{solar_nebula_value}, \citet{rosetta_n2_co_1}; 2. This Work/\citep{kumar_epsc,Kumar_thesis};  3. \citet{n2_co_Biver}; 4. \citet{cochran_co_c2016r2, cochran_co_c2016r2_erratum};  5. \citet{n2_co_opitom}; 6. \citet{n2_co_Mckay}; 7. \citet{n2+_comets_1};  8. \citet{rosetta_n2_co_1}; 9. \citet{hale-bopp-cochran}}
\begin{tabular}{|c | c | c | c | c | c | c | }
\hline
\textbf{Object} \Tstrut\Bstrut & \textbf{r} & \textbf{$\Delta$} & \textbf{Instrument} & \textbf{Observatory} &  \textbf{N$_{2}$/CO} & \textbf{Reference}   \\
\TTstrut\BBstrut     & (AU) & (AU) &  &   \\ \hline
\Tstrut\Bstrut Solar Nebula & -- & -- & -- & -- & 0.145  & 1  \\
\hline
\Tstrut\Bstrut C/2016 R2 &  2.8   & 2.1 & LISA (R=1000) & 1.2m Mount Abu & 0.09 $\pm$ 0.02 & 2   \\
\hline
\Tstrut\Bstrut C/2016 R2 &  2.9  & 2.0 & LISA (R=1000) & 0.28m telescope & 0.09 $\pm$ 0.03 & 3  \\ \hline
\Tstrut\Bstrut C/2016 R2 &  2.7   & 2.5 & Alpy600 (R=500) & 0.2m telescope  & 0.06 $\pm$ 0.02 & 3 \\ \hline
\Tstrut\Bstrut C/2016 R2 &  3.0   & 2.1 & Tull 2DCoude (R=60000) & 2.7m Mcdonald Observatory  & 0.06 & 4 \\ \hline
\Tstrut\Bstrut C/2016 R2 &  2.8   & 2.4 & UVES (R=80000) & 8.2m VLT & 0.06 $\pm$ 0.01 & 5\\ \hline
\Tstrut\Bstrut C/2016 R2 &  2.8   & 2.3 & ARCES (R=31500) & 3.5m Apache Point Observatory & 0.05 $\pm$ 0.01 & 6\\ \hline
\Tstrut\Bstrut C/2002 VQ94 &  6.8 & 5.7 & SCORPIO & 6m telescope BTA & 0.04 &  7\\ \hline
\Tstrut\Bstrut 67P/C-G &  3.1  & 10(km)$^{\dagger}$ & ROSINA & Rosetta & 5.70 $\pm$ 0.66 $\times 10^{-3}$  & 8 \\ \hline
\Tstrut\Bstrut 122P/deVico &  0.6 & 1.0 & Tull 2DCoude (R=60000) & 2.7m Mcdonald Observatory & 3 $\times 10^{-4}$  & 9  \\ \hline
\Tstrut\Bstrut C/1995 O1 &  0.9   & 1.4 & Tull 2DCoude (R=60000) & 2.7m Mcdonald Observatory & 9.9 $\times 10^{-5}$  & 9 \\ 
\hline \\

\end{tabular}
\label{n2_co_ratios}
\vspace*{0.1cm}
\small{\\ $^{\dagger}$ Rosetta observed the comet 67P/C-G at a distance of 10km from the nucleus in Oct 2014. Observations carried out from May to Oct 2014.}
\end{table*}

\subsection{Search for possible emissions from NH$_{2}$ and H$_{2}$O$^{+}$}
\label{line_id}
With the high resolution spectroscopic observations on UVES-VLT, \citet{n2_co_opitom} have detected various emission lines with strong signatures from CO+, CO$_{2}$+ and N$_{2}$+. However they did not detect any emissions from OH, OH$^{+}$, NH$_{2}$ and H$_{2}$O$^{+}$. Our observations were made about one month prior to their UVES-VLT observations, when the comet was at a relatively lower phase angle. 
We are limited by our spectral resolution. However, in order to understand some of the minor emission features seen on the redder side of 5500 \AA~ in our spectra, we have made an attempt to compare the emission features with those from different catalogues and laboratory spectra. Our conclusions are based on how these features matched with various cometary spectral features listed in the catalogues. However, their presence seems to be ambiguous owing to various blends in the low resolution spectra. These are clarified with the high resolution spectra of the comet \citep{n2_co_opitom}. It should be noted, though, that they 
have mentioned many of the lines in these regions as 'unidentified'. The peculiar composition of this comet has also been studied and reported by \citet{n2_co_Biver}, using the 30m IRAM telescope. They have also reported complimentary observations with LISA spectrograph on the 0.28m telescope. However, they have not reported the spectra beyond 550 nm owing to low signal-to-noise ratio in their spectra.

There have been a few comets in the past whose emission lines have been catalogued based on their high resolution optical spectrum. \citet{cometary_lines} have catalogued emission lines from comet 109P/ (Swift-Tuttle) and comet Brorsen-Metcalf (23P/). \citet{comet_lines_cochran}, \citet{comet_lines_zhang} and \citet{comet_lines_153p} have catalogued the emission lines of comets 122P/(de Vico), C/1995 O1 (Hale-Bopp) and 153P/(Ikeya-Zhang) respectively, based on their high resolution spectra. 
Emission lines from different molecular and ionic species from each of these catalogues were used to generate a synthetic\footnote{The phrase 'synthetic spectra' usually refers to spectra generated using theoretical calculations. However, we use this phrase for the low resolution spectrum generated using the high resolution cometary lines from the catalogue} spectrum, which matches the resolution of the observed spectra of comet R2. Each emission line in these catalogues was represented as a gaussian profile with a width matching the line profile width in the observed spectrum. All of these gaussian profiles were then summed up as a function of wavelength. The result was a spectrum with a resolution similar to that of our observed spectrum. The synthetic spectrum generated using all of the above catalogues were matched with our observed spectrum. We note that, not all the lines are present with the same relative intensities in all the comets with available high-resolution spectra. It was seen that the spectrum of comet 109P/ (Swift-Tuttle) forms the best match for our observed spectrum of comet R2. Therefore, this spectrum of the comet 109P/ by \citet{cometary_lines} was used to identify the emission bands seen in our observed spectrum. In the low resolution spectra, the blending of various emission lines define the pattern of the observed emission band. Therefore, the identification was based on wavelength coincidence and by visually matching the emission band patterns in the observed spectrum with that of the synthesized one. Such an initial analysis indicated a possible presence of emissions from NH$_{2}$ and H$_{2}$O$^{+}$ in the spectrum.
In order to further investigate our results and search for the presence of these species, we have used the CO$^{+}$ laboratory spectra \citep[][and personal communication with Dr.Ivanova, Astronomical Institute of the Slovak Academy of Sciences]{CO_plus_lab} to compare the emission features seen in the observed spectrum. There are certain features of H$_{2}$O$^{+}$ and NH$_{2}$ which blend completely with some of the CO$^{+}$ emissions. However, there are also certain emission features which stand out and  which do not blend with the CO$^{+}$ emission bands. This is explained in more detail in the sub-sections below.
\begin{figure*}
\centering
  \includegraphics[width=0.8\textwidth]{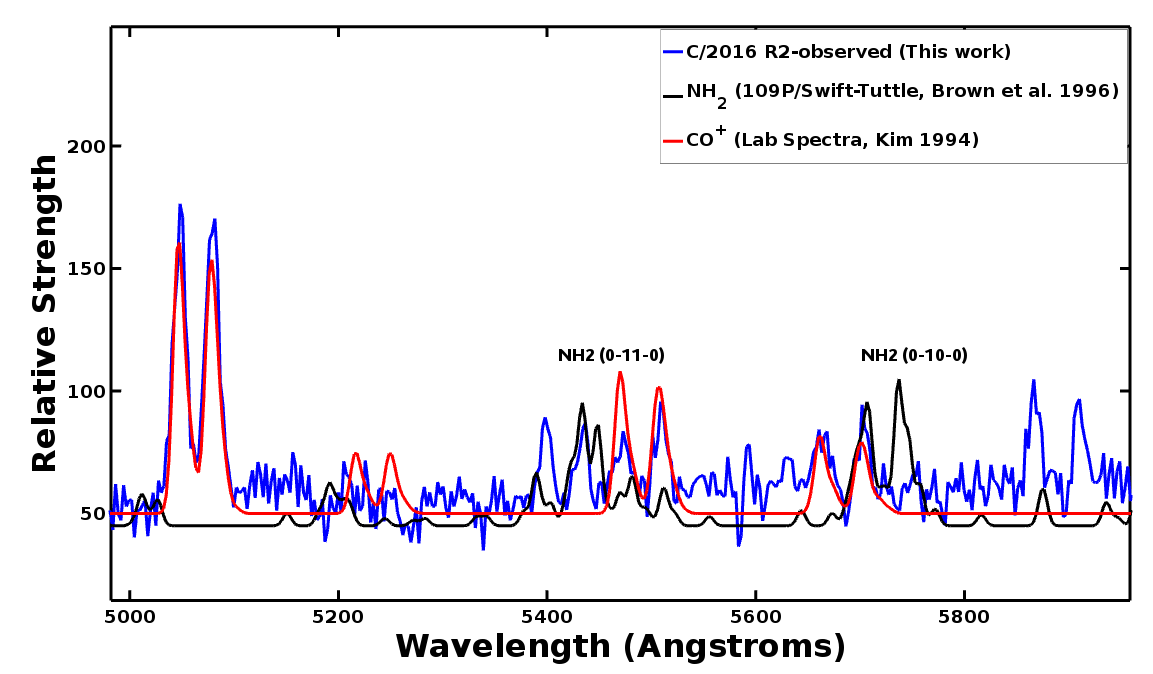}
 \caption{The observed spectrum of comet C/2016 R2 (blue) plotted along with the synthetic spectra generated using the catalogued NH$_{2}$ lines (black) in comet 109P/Swift-Tuttle \citep{cometary_lines}. The red spectrum is the CO$^{+}$ laboratory spectra \citep{CO_plus_lab}. The NH$_{2}$ (0-10-0) band is strongly blended with a CO$^{+}$ feature. The NH$_{2}$ (0-11-0) emission feature is relatively better resolved. However there is a minor blend with the (0-1) CO$^{+}$ doublet towards the edge of the profile.} 
 \label{nh2_spectra}
\end{figure*}

\subsubsection*{Search for NH$_{2}$ emission features}
The figure \ref{nh2_spectra} shows the observed spectrum plotted along with the NH$_{2}$ synthetic spectrum \citep{cometary_lines} and the CO$^{+}$ lab spectrum \citep{CO_plus_lab}. Two strong emission band features in the observed spectrum significantly matched the NH$_{2}$ synthetic spectrum. One of them is the NH$_{2}$ (0-10-0) band near 5700 \AA, and second one is the (0-11-0) band near 5430 \AA. The (0-10-0) band coincides with a CO$^{+}$ doublet at 5700 \AA. The (0-11-0) band is relatively less contaminated and better resolved. There is a minor blend towards the end of its profile from the nearby CO$^{+}$ (0-1) doublet emission feature centred around 5460 \AA~ and 5500 \AA. A few other minor features were also detected, which could be attributed to NH$_{2}$ emissions e.g (1-7-0) band near 5380 \AA. However, strong emissions from (0-9-0) band centred around 6000 \AA~ are missing in the observed spectrum. 
The column densities and production rates of 0-11-0 emission bands were calculated with $3\sigma$ (photon-noise of 10\%) upper limits. These are listed in table \ref{nh2_production}. The g-factors from \citet{nh2_g_factor} and daughter scale-lengths from \citet{nh2_scalelength_fink} were used in the process. We assume that NH$_{3}$ is the only parent of NH$_{2}$ which is produced through the photolytic process \citep{nh3_to_nh2_wyckoff_1} NH$_{3}$ + h$\nu$ $\longrightarrow$ NH$_{2}$ + H. Since this process has a branching ratio of 95\%, it would be safe to assume that the production rate of NH$_{3}$ will be equivalent to that obtained from the NH$_{2}$ band fluxes. 
A synthetic spectrum was generated (as described in section \ref{line_id}) using the strengths of the lines in NH$_{2}$ bands given in \citet{cometary_lines} in order to examine the blending of different emission bands in the observed low resolution spectrum. The 0-11-0 band was contaminated by the presence of 1-7-0 band. The intensity ratio of 1-7-0 to 0-11-0 emissions within the selected wavelength region was calculated based on the strengths given by \citet{cometary_lines}. 
This ratio was then used to remove the flux contribution of 1-7-0 emission in the 0-11-0 band of the observed comet spectrum. The 0-11-0 band is also contaminated at the tail end of its profile by $\Pi_{1/2}$ emission of the (0-1) band of CO$^{+}$ which is centred at 5461. The relative intensity of this band is unknown and has also not been explicitly calculated by \citet{co_theory}. It is therefore difficult to quantify the amount of contamination in the (0-11-0) band of NH$_{2}$. Since we are overestimating the (0-11-0) band flux, the calculated column densities and production rates for this band represent an upper limit.

\begin{figure*}
\centering
 \includegraphics[width=0.8\textwidth]{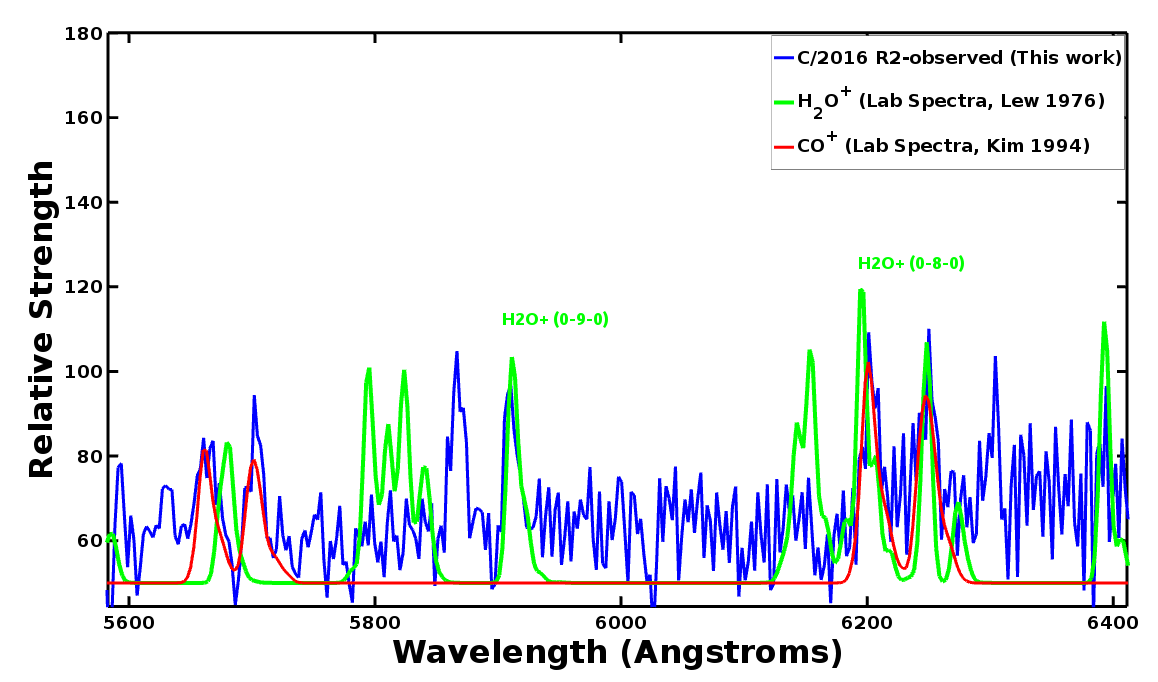}
 \caption{The observed spectrum of comet C/2016 R2 (blue) plotted along with the synthetic spectra generated using the laboratory spectra \citep{lew_h2o+} of H$_{2}$O$^{+}$ lines (green). The CO$^{+}$ (red) laboratory spectra \citep{CO_plus_lab} is shown for comparison. The (0-8-0) band at 6200 \AA blends strongly with the CO$^{+}$ feature. However, the (0-9-0) feature at 5910 \AA is seen without any blend.}
 \label{h2o+_spectra_lab_co}
\end{figure*}

\begin{figure*}
\centering
 \includegraphics[width=0.8\textwidth]{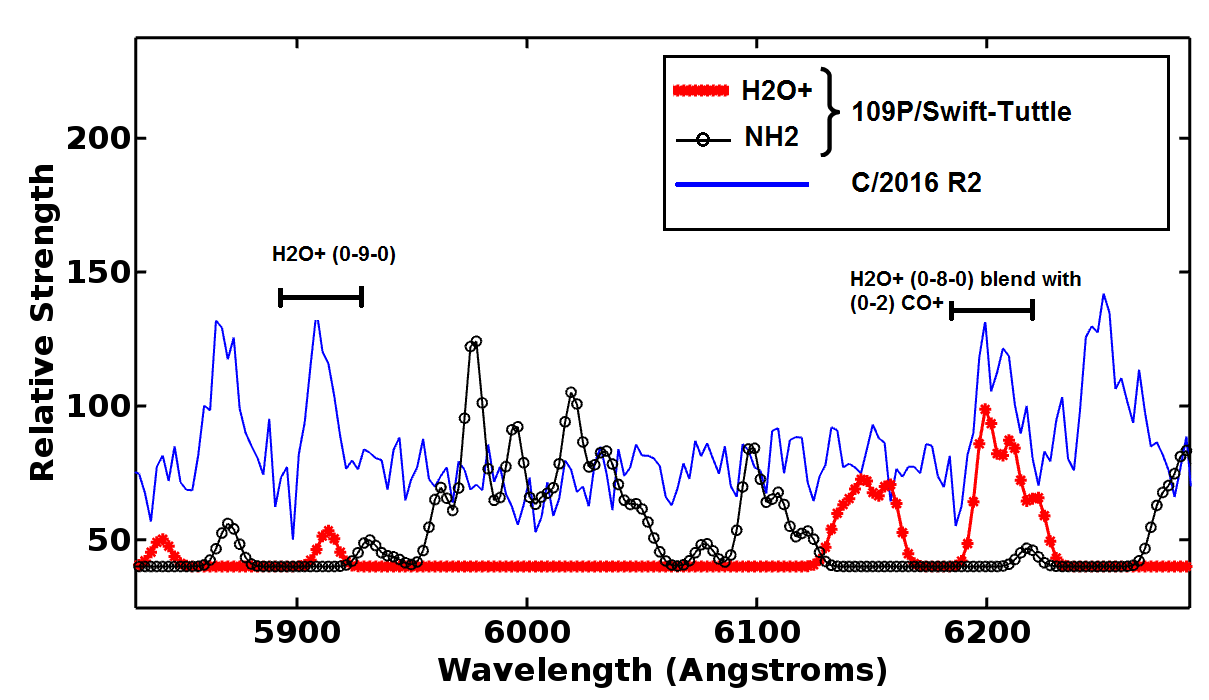}
 \caption{The observed spectrum of comet C/2016 R2 plotted along with the synthetic spectrum generated using the catalogued H$_{2}$O$^{+}$ lines in comet 109P/Swift-Tuttle \citep{cometary_lines}. The NH$_{2}$ synthetic spectrum has also been plotted in the figure to check for any kind of blend between the emissions from both the species.}
 \label{h2o+_spectra}
\end{figure*}

\begin{figure*}
\centering
 \includegraphics[width=0.8\textwidth]{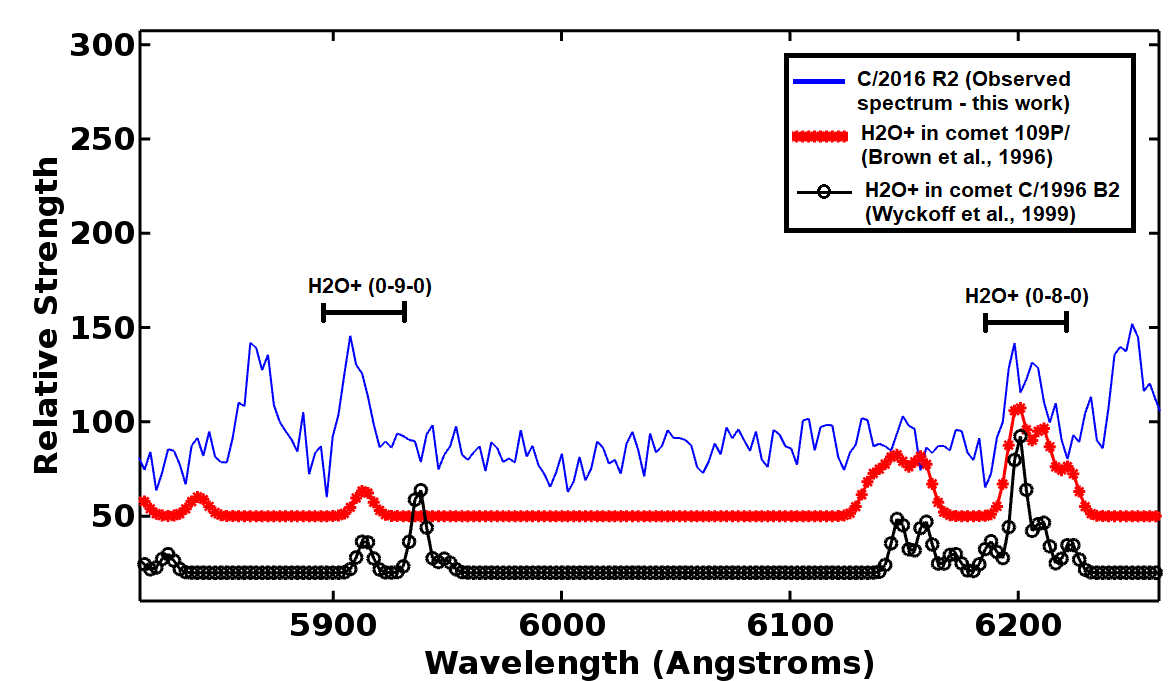}
 \caption{The observed spectrum of comet C/2016 R2 plotted along with the synthetic spectra generated using the catalogued H$_{2}$O$^{+}$ lines in comet 109P/Swift-Tuttle \citep{cometary_lines} and H$_{2}$O$^{+}$ lines in comet C/1996 B2 \citep{hyakutake_1}}
 \label{h2o+_spectra_wyckoff}
\end{figure*}

Prior to comet R2, the only comet in which N$_{2}^{+}$, CO$^{+}$ and NH$_{2}$ have been detected simultaneously is comet Halley. \citet{n2_nh3_ratio} have determined the average N$_{2}$/NH$_{3}$ ratio in comet Halley and in various star-forming regions. They report a value of 0.1 for this ratio in comet Halley which differs significantly from the values for the star forming regions. All of these three species have been detected in comet R2, although the detection of NH$_{2}$ seems to be only marginal.

\subsubsection*{Search for H$_{2}$O$^{+}$ emission features}
Figure \ref{h2o+_spectra_lab_co} shows the observed spectrum plotted along with H$_{2}$O$^{+}$ \citep{lew_h2o+} and CO$^{+}$ \citep{CO_plus_lab} lab spectra. The lab spectra are scaled to relative strengths in the vertical axis in order to look for matching features. The (0-8-0) H$_{2}$O$^{+}$ emission feature centred around 6200 \AA~and the (0-9-0) feature centred at around 5910 \AA~in the lab spectrum matches well with the observed spectrum. The (0-8-0) is severely contaminated by the CO$^{+}$ (0-2) doublet and therefore, their individual contributions cannot be determined for (0-8-0) H$_{2}$O$^{+}$ and (0-2) CO$^{+}$ band at 6200 \AA. However, the (0-9-0) H$_{2}$O$^{+}$  feature does not have any contamination from the CO$^{+}$ emissions. \\ 
The observed spectrum was also compared to the H$_{2}$O$^{+}$ emissions from other comet observations. It was seen, that both the (0-8-0) and (0-9-0) emission features matched quite well with the corresponding H$_{2}$O$^{+}$ emission features listed in the catalogues for other comets. This can be seen in figures \ref{h2o+_spectra}  and \ref{h2o+_spectra_wyckoff}, where the spectrum of comet R2 is compared with the H$_{2}$O$^{+}$ synthetic spectra created by degrading the high resolution spectra of comets  109P/ (Swift-Tuttle) and C/1996 B2 (Hyakutake) observed by \citep{cometary_lines} and \citet{hyakutake_1} respectively. The NH$_{2}$ synthetic spectrum has also been plotted, in the figure, to check for possible blend between the emissions from the two species in this wavelength region. Though there is a very weak blend of the (0-9-0) feature at 5910 \AA~with one NH$_{2}$ band at 5931 \AA~(the blend occurs at the tail of the line profile), the two features can be significantly resolved in the observed spectrum. \\ 
Based on the above analysis, the emission feature at 5910 \AA~appears to be a relatively better indicator for the presence of H$_{2}$O$^{+}$ ions. As seen from figure \ref{h2o+_spectra_lab_co}, the emission feature centred around 6200 \AA~is expected to have a strong blend with the $\Pi_{1/2}$ (0-2) emission of CO$^{+}$ at 6189 \AA. Owing to the low resolution of the observed spectrum, it is not possible to estimate the extent of contamination using flux ratios with the stronger CO$^{+}$ bands. However, this blend and the match with the synthetic spectra (figure \ref{h2o+_spectra}) can be understood in the following way. 

\begin{table*}
\centering
\caption{ NH$_{2}$ emission in comet C/2016 R2:.The $3\sigma$ (photon noise of 10\%) upper limits of Column Density and Production rate have been listed.}
\begin{tabular}{|c | c | c | c | c | c | c |}
\hline
Band Transition \Tstrut\Bstrut & Wavelength range & g-factor & \multicolumn{2}{c|}{Column Density} & \multicolumn{2}{c|}{Production rate}  \\
     &  (\AA)      &  ($\times$ 10$^{-3}$ photons mol$^{-1}$ sec$^{-1}$) & \multicolumn{2}{c|}{($\times$ 10$^{11}$ mol cm$^{-2}$)} &
    \multicolumn{2}{c|}{ ($\times$ 10$^{23}$ mol sec$^{-1}$)} \\
\cline{4-7}
      & & & 13/01/2018 \Tstrut\Bstrut & 25/01/2018 & 13/01/2018 \Tstrut\Bstrut & 25/01/2018  \\    \cline{1-7}
(0-11-0) \Tstrut\Bstrut & 5410-5495 & 4.17 & $<$ 3.40 & $<$ 3.15 & $<$ 4.82 & $<$ 5.91 \\ \hline

\end{tabular}
\label{nh2_production}
\end{table*}


The H$_{2}$O$^{+}$ synthetic spectrum has been generated using the catalogued \citep{cometary_lines} high resolution H$_{2}$O$^{+}$ emission lines of comet 109P/Swift-Tuttle. The same comet has been imaged by \citet{comet_swift_co} in CO$^{+}$ (2-0) band and H$_{2}$O$^{+}$ (0-8-0) band filters centred around 4260 \AA~ and 6200 \AA~ respectively. Using the g-factors from \citet{co_theory}, they claim that the H$_{2}$O$^{+}$ filter is contaminated by CO$^{+}$ (0-2) band up to an extent of $\approx$ 17\%. Both \citet{cometary_lines} and \citet{comet_swift_co} have observed the comet within a gap of 7 days and in both the cases, the comet was at a heliocentric distance of about 1 AU. \citet{comet_swift_co} have quoted ions per unit tail length ratios of CO$^{+}$/H$_{2}$O$^{+}$, in the tail-ward direction at distances of 10$^{5}$ km, to be $\approx$ 0.5. However, due to contamination close to the nucleus, they do not quote any value for the total ion content. Even if we consider  CO$^{+}$ to be diffused close to the nucleus, looking at figure 10 of \citet{comet_swift_co}, the CO$^{+}$/H$_{2}$O$^{+}$ ratio varies from 50 to 60\% at the nuclear centre, which could be a significant value. 
As discussed above, the H$_{2}$O$^{+}$ synthetic spectra created based on the spectral catalogue in \citet{cometary_lines} matches quite well with the emission feature around 6200 \AA~ in the observed spectrum of R2. \citet{cometary_lines} obtained the high resolution spectra by centring the comet image on the slit, which implies, that they are looking at the photo-centre of the comet. Though they have catalogued the H$_{2}$O$^{+}$ lines, there is no mention or indication of CO$^{+}$ emissions in their spectra. They have identified the H$_{2}$O$^{+}$ emission lines based on the laboratory spectra given by \citet{lew_h2o+}. 
Therefore, it is possible that the catalogued high resolution emission lines of H$_{2}$O$^{+}$ might be contaminated with CO$^{+}$ emissions and subsequently, the synthetic spectra derived from this catalogue might also have blended emissions from both CO$^{+}$ and H$_{2}$O$^{+}$, thereby matching well with the observed spectra. 
\\ 
It is difficult to conclude the presence of H$_{2}$O$^{+}$ based on the 6200 \AA~ features because of the blend with the CO$^{+}$ doublet. The emission at 5910 \AA~ stands out as a separate feature free of any contamination from CO$^{+}$. The integrated flux ratio of this emission feature with respect to the CO$^{+}$ (3-0) doublet is 0.18 $\pm$ 0.04. \citet{n2_co_Mckay} have measured the H$_{2}$O/CO ratio to be 0.32\%, which is significantly lower than the flux ratio that we have measured. However, since the g-factors for the H$_{2}$O$^{+}$ (0-9-0) is unknown, the conversion of the flux ratio into ratio of production rates of H$_{2}$O$^{+}$/CO$^{+}$ is not feasible. Although it does not conclusively prove the  detection of emissions from H$_{2}$O$^{+}$ ions, above evidences suggest that their presence in comet R2 is highly probable.

\section{Discussion}
\label{discussion}
The unusual spectrum of comet R2 raises a lot of questions on the comet's origin and its formation in the solar system. The implications of such a spectrum and the likely origin of the comet has been discussed by \citet{Kumar_thesis} at length. The detection of N$_{2}^{+}$ or the parent N$_{2}$ in comets is vital, as they provide key information on the formation of ices in the early solar nebula and their agglomeration in comets. The abundance of N$_{2}$ in a comet will strongly depend on the location (and hence temperature), where the comet was formed. This would define the process of comet agglomerating the condensed ices from the nebula, trapping the gases and the chemistry thenceforth. Since N$_{2}$ and CO condense and evolve at similar temperatures \citep{n2_co_condense}, measuring the N$_{2}$/CO ratio becomes vital. The N$_{2}$/CO ratio calculated for comet R2 (see table \ref{n2_co_ratios} for all references) is the highest value to be reported as compared to any of the other comets in which this ratio has been measured. The comparison of this ratio with some of the other comets and its value in the solar nebula  \citep[as calculated by][in the gas phase proto-solar Nebula]{rosetta_n2_co_1} is shown in figure \ref{ratio}. \citet{hale-bopp-cochran} determined the N$_{2}^{+}$/CO$^{+}$ ratios in comets 122P/de Vico and comet Hale-Bopp and  derived the upper limits as 3 $\times$ 10$^{-4}$ and 9.9 $\times$ 10$^{-5}$ respectively. They have also referenced many other works in which N$_{2}^{+}$ detection has been reported and the N$_{2}^{+}$/CO$^{+}$ ratio has been measured. \citet{n2+_comets_1} have detected N$_{2}^{+}$ in comet C/2002 VQ94 at a heliocentric distance of 6.8 AU and have reported a significantly larger value of N$_{2}^{+}$/CO$^{+}$ = 0.04. However, they claim N$_{2}^{+}$ as only a tentative detection. In general, the N$_{2}$/CO ratios measured in comets seem to be significantly depleted as compared to the ratios of the proto-solar nebula (PSN) itself. \citet{n2_depletion} have pointed out, that the deficiency of molecular nitrogen as compared to CO is because of the fact that CO forms clathrate hydrate more easily than N$_{2}$. This also depends on the amount of water ice available for the clathration  to take place.   
One of the recent and probably the most authentic in-situ measurements of N$_{2}$/CO ratio in comet 67P/CG was made by ROSINA mass spectrometer on board the Rosetta spacecraft\citep{rosetta_n2_co_1}. They report an average value for the ratio N$_{2}$/CO = 5.70 $\times$ 10$^{-3}$. Assuming a proto-solar gas phase value of N$_{2}$/CO = 0.145, they have estimated the depletion factor of this ratio to be $\approx$ 25.4. They have interpreted this depletion as the inefficient trapping of N$_{2}$ in amorphous water ice or the clathrate cages formed by crystallised water ice. However, they also quote, that this ratio might increase below a temperature of 24 K due to increased efficiency of N$_{2}$ trapping.
The measured value of N$_{2}$/CO ratio in comet R2 is larger than the mean value of comet 67P/ by about an order of magnitude. Assuming an N$_{2}$/CO ratio of 0.145 for the solar nebula, the depletion factor based on our calculated value of this ratio turns out to be about $\approx$ 1.6 $\pm$ 0.4 This raises the question as to why this comet has a much smaller depletion factor as compared to the others. The answer is again related to the formation of the comet's ices.
\citet{cochran_co_c2016r2, cochran_co_c2016r2_erratum} have reported that the ratio in comet R2 (N$_{2}$/CO = 0.06) is consistent with the laboratory experiment results \citet{Owen_bar_nun_1995}, where a similar ratio is expected in the gases trapped in amorphous water ice, assuming that the icy planetesimals formed in the solar nebula at around 50K and N$_{2}$/CO $\approx$ 1 in the nebula itself. More recently, the same group \citep{Bar_nun_2007} had carried out laboratory experiments where the rate of ice deposition was significantly diminished to 0.1-10$^{-5}$ $\mu$m min$^{-1}$ reaching an ice thickness of 0.1 $\mu$m as compared to a faster rate of 0.3 $\mu$m min$^{-1}$ reaching an ice thickness of 10 $\mu$m in the early experiments (Lofer 2018, priv. comm). They quote, that the abundance of trapped gases would highly depend on ice deposition rates and have found that the rapid deposition may be inappropriate for the conditions in the interstellar medium or the PSN. With the slower deposition rate and assuming a nebular N$_{2}$/CO value of 0.22  \citet{Bar_nun_2007} have estimated N$_{2}$/CO ratio in comets to be 6.6 $\times$ 10$^{-3}$ using the experimentally found depletion factor\footnote{This depletion factor corresponds to a value of 33.3 in terms of the same factor calculated by \citet{rosetta_n2_co_1}, which is an inverse as compared to the one calculated by \citet{Bar_nun_2007}.} of 3 $\times$ 10$^{-2}$. This is much smaller than the value obtained for comet R2 and is not consistent with the model of N$_{2}$ being trapped by amorphous water ice. \citet{rosetta_n2_co_1} and \citet{rosetta_n2_co_2} have both cited \citet{Bar_nun_2007} in order to compare and interpret N$_{2}$/CO ratio of comet 67P. 
We refer to the work of \citet{Bar_nun_2007}, \citet{rosetta_n2_co_1} and \citet{rosetta_n2_co_2} in order to explain the large N$_{2}$/CO ratio found in comet R2. There are two possible scenarios:
\begin{enumerate}
 \item \textbf{N$_{2}$ trapped in amorphous water ice:} If N$_{2}$ in comet R2 was trapped by amorphous water ice, the trapping efficiency has to be much greater than what has been observed in other comets and this would be possible if the temperature is much lower than 24K \citep{rosetta_n2_co_1}. At 24K, the depletion factor is found to be about $\approx$ 19, whereas, the depletion factor for comet R2 is found to be $\approx$ 1.6.
 \item \textbf{Condensation of N$_{2}$ and CO as pure ices:} Figure 1 of \citet{rosetta_n2_co_2} gives a detailed comparison of N$_{2}$/CO ratio measured in comet 67P/ to that of the one sampled from the (i) gases trapped in amorphous ice, (ii) case where N$_{2}$ and CO are crystallised as pure ices in the proto-solar nebula and (iii) as gases trapped in the clathrate cages. The ratio measured for comet 67P is consistent with the laboratory experiment results of N$_{2}$ being trapped in amorphous ice at around 24-30 K range. If we consider trapping in clathrate cages, the 67P/ data is consistent with this result in the temperature range of 44-50 K. The higher N$_{2}$/CO ratio for comet R2 does not seem to match both the above mentioned results. However, it does match the results in the case where N$_{2}$ and CO are crystallised as pure ices. This is possible when there is scarcity of water ice to either form clathrate hydrates or amorphous ice to trap these gases.   
\end{enumerate}

The implications of both the above scenarios would be quite different. We refer to the introductory section of \citet{rosetta_n2_co_2} and references therein to understand the formation of two reservoirs of ices in the proto-solar nebula (PSN). In the proto-solar nebula, the icy planetesimals within 30 AU of Sun, which agglomerated ices from the interstellar medium were vaporised due to their solar vicinity. In the process of cooling of the proto-solar nebula, water condensed at around 150 K to form crystalline ice. As long as crystalline water was available, the planetesimals formed in this region agglomerated ices in the form of clathrates. In the case where water was not sufficient or clathration was not possible, pure CO and N$_{2}$ ices would have possibly formed. If we assume that the second scenario is true in case of comet R2, there is definitive probability that this Oort cloud comet formed and originated in the inner regions of the solar system, within 30 AU of Sun. This is not surprising, as there have been many theories and models, which predict that the Oort cloud comets initially originated very close to the sun and were then propelled into the Oort cloud by the gravitational influence of giant planets e.g. \citep{oort_1, oort_2}.
In case of the first scenario, the comet must have formed at very large heliocentric distance, either at the edge of the PSN, or in the outer interstellar medium, which is then captured by the Oort cloud.
However, in case of both the formation scenarios mentioned above, the unusual spectrum of the comet is still not justified. Not only does comet R2 come under the category of carbon poor comets (absence of C$_{2}$ and C$_{3}$ emissions, but not accounting for the Carbon in CO), but also the absence of the cyanogen emission and presence of NH$_{2}$ once again poses intriguing questions on the formation and origin of the comet. If we ignore the two major ionic emissions in this comet, it becomes quite similar to the unusual comet 1988r (Yanaka) which was studied in detail by \citet{fink_yanaka}. This comet showed a lot of NH$_{2}$ emissions and also possibly CO emissions (not confirmed, but reported in the article), but no signatures of major emissions from C$_{2}$ or CN. \citet{fink_yanaka} explains the possible origin of this comet as coming from the outer molecular clouds with a composition considerably different from that of the known ones. Studies on the recently seen interstellar comet 2I/Borisov \citep{borisov_opitom, borisov_4} show that the prediction by \citet{fink_yanaka} could, indeed, be true. In fact interstellar comets may be more common than they appear to be \citep{borisov_2, borisov_3}. A similar conclusion could be drawn for comet R2 due to its unusual composition questioning the location of its formation. The study by \citet{2019EPSC...13.1136M_mousis} also explains the condensation scenarios of comet R2 based on the recent observations. 

\section{Conclusion}
Our spectroscopic observations of the comet C/2016 R2 from the Mount Abu Infra-red Observatory have revealed its unusual behavior. The comet spectra did not exhibit any of the major cometary emissions from the neutral species (except NH$_{2}$), which are expected in an Oort cloud comet at the observed heliocentric distance. The following conclusions are drawn from the present spectroscopic observations of the comet : 
\begin{enumerate}
\item The optical spectrum is strongly dominated by emissions from CO$^{+}$ and N$_{2}^{+}$. 
\item The comet spectrum also exhibited emission features on the redder side of 5400 \AA~. A search for possible emissions from H$_{2}$O$^{+}$ and NH$_{2}$ were carried out by comparing the observed spectrum with the CO$^{+}$ lab spectrum and spectral features from other comets. The (0-9-0) H$_{2}$O$^{+}$ and (0-11-0) NH$_{2}$ emissions matched the observed spectra without any contamination from CO$^{+}$ emissions. 
\item The N$_{2}$/CO ratio estimated for this comet ($0.09\pm0.02$) is highest ever reported for any comet for which such ratio has been measured. The ratio is found to be uniform, within the measurement errors, over a distance of $\sim$18000 km from photo-centre.
\item The extremely small depletion factor of N$_{2}$/CO ratio for comet R2 could be an indication of chemically different conditions that persisted at the location of formation of the comet in the proto-solar nebula.
\item Two possible formation scenarios have been proposed in which either N$_{2}$ is trapped in amorphous water ice at low temperatures below 24K or it has condensed as pure ice. 
\end{enumerate}

\section*{Acknowledgments}
\small This work is supported by the Dept. of Space, Govt. of India. We thank the referee, Dr Michael DiSanti, for critically reviewing the manuscript and providing comments which have improved this paper.
We would like to specially thank Dr. S. Raghuram (PRL) and K. Aravind (PRL) for useful discussions. We acknowledge the local staff at the Mount Abu Infra-Red Observatory for their help and a special thanks to Mr. Prashant Chauhan \& Mr. Jinesh Jain, for their assistance in the observations. We would like to thank Dr. Diana Lofer and Prof. Dennis Bodewits for their invaluable inputs and suggestions. We would like to specially thank Dr. Oleksandra Ivanova for providing the lab spectra for our analysis. We also thank our colleagues in the Astronomy \& Astrophysics division at PRL for their comments and suggestions. \\
This research has made use of ephemerides from NASA HORIZONS system and NASA's Astrophysics Data System Bibliographic Services

\bibliographystyle{mnras}
\bibliography{references}

\pagebreak

\onecolumn

\begin{appendix}

 \section{Production Rates and Column Densities}
 The production rate of NH$_{2}$ have been derived using the following formula \citep{comets_book_ks}
 
\begin{equation}
Q=\frac{4\pi \Delta^{2}}{g \ \tau} \times Flux 
\label{Q_eq}
\end{equation}
where $\Delta$ is the earth-comet distance, $g$ is the fluorescence efficiency of the molecule and $\tau$ is the lifetime of the molecule.

The observed column density is calculated as given below.

\begin{equation}
 N=\frac{4 \pi}{g} \frac{F}{\Omega}
 \label{CD_1}
\end{equation}
where $g$ is the fluorescence efficiency in ergs per molecule per second, and $\Omega$ is the solid angle subtended by the aperture in steradians. $\Omega$ is calculated as the product of the slit width and size of the aperture used to extract the spectrum.

The calculated column densities of the CO$^{+}$, NH$_{2}$ and N$_{2}^{+}$ in comet C/2016 R2 have been tabulated in tables \ref{co_production}, \ref{nh2_production} and \ref{n2_production} respectively. 

\subsection*{}
The column density ratios are calculated as 

\begin{equation}
 \frac{N(N_{2}^{+})}{N(CO^{+})}= \frac{F(N_{2}^{+})g(CO^{+})}{g(N_{2}^{+})F(CO^{+})}
\end{equation}

The calculated column density ratios in comet C/2016 R2 have been tabulated in table \ref{n2_production}. 

\section*{Error Estimation}
The errors in the calculations have been estimated using the error formula (photon noise) $\sqrt{n}$, where $n$ is the number counts in the raw spectrum (Assuming a Poisson distribution for a large number of photons). This results in $\approx$ 10\% errors in the calculated fluxes, column densities and production rates.

\end{appendix}

\label{lastpage}
\end{document}